# Magic silicon dioxide for widely tunable integrated photonics

Bruno Lopez-Rodriguez[1*†], Naresh Sharma[1†], Zizheng Li[1], Roald van der Kolk[1], Jasper van der Boom[1], Thomas Scholte[1], Jin Chang[2], Simon Gröblacher[2], Iman Esmaeil Zadeh[1]

[1]Department of Imaging Physics (ImPhys), Faculty of Applied Sciences, Delft University of Technology, Lorentzweg 1, Delft, 2628 CJ, The Netherlands.

[2]Department of Quantum Nanoscience, Faculty of Applied Sciences, Delft University of Technology, Lorentzweg 1, Delft, 2628 CJ, The Netherlands.

*Corresponding author(s). E-mail(s): b.lopezrodriguez@tudelft.nl;

Contributing authors: N.Sharma-1@tudelft.nl;

†These authors contributed equally to this work.

**Abstract**

Integrated photonic circuits have transformed data communication, biosensing, and light detection and ranging, and hold wide-ranging potential for optical computing, optical imaging and signal processing. These applications often require tunable and reconfigurable photonic components, most commonly accomplished through the thermo-optic effect. However, the resulting tuning window is limited for standard optical materials such as silicon dioxide and silicon nitride. Most importantly, bidirectional thermal tuning on a single platform has not been realized. For the first time, we show that by tuning and optimizing the deposition conditions in inductively-coupled plasma chemical vapor deposition (ICPCVD) of silicon dioxide, this material can be used to deterministically tune the thermo-optic properties of optical devices without introducing significant losses. We demonstrate that we can deterministically integrate positive and negative wavelength shifts on a single chip, validated on amorphous silicon carbide (a-SiC), silicon nitride (SiN) and silicon-on-insulator (SOI) platforms. We observe up to a 10-fold improvement of the thermo-optic tunability and, in addition, demonstrate athermal ring resonators with shifts as low as $1.5\,\mathrm{pm/°C}$. This enables the fabrication of a novel tunable coupled ring optical waveguide (CROW) requiring only a single heater. In addition, the low-temperature deposition of our silicon dioxide cladding can be combined with lift-off to isolate the optical devices resulting in a decrease in thermal crosstalk by at least two orders of magnitude. Our method paves the way for novel photonic architectures incorporating bidirectional thermo-optic tunability.

# 1 Introduction

Achieving a high degree of tunability in photonic devices has been a focal point in the field of integrated photonics for several decades with a wide range of applications from telecommunications and biochemical sensing to fundamental quantum photonic experiments in many material platforms[1–17].

The most universally utilized method to achieve tunability in photonic devices is by exploiting the thermo-optic effect. The thermo-optic coefficient (TOC) of an optical material describes the change in refractive index due to a temperature change (dn/dT)[18–20]. In the usual configuration, metal heaters are placed above the guiding material to control the phase of the light. This method of tuning photonic devices is virtually lossless, easy to integrate, and applicable to nearly all photonic platforms. Nevertheless, the tuning strength is specific to the material platform and, importantly, it is weak in photonic platforms such as silicon nitride or silicon dioxide, two of the most commonly used materials in integrated photonics [21, 22]. Increasing the TOC of materials has been a major challenge and, so far, accomplished by tuning their composition [23, 24] or depositing high refractive index claddings [25]. However, both methods are complex and only applicable to specific platforms. Moreover, changing the composition modifies the overall properties of the guiding layer and often significantly increases the propagation losses while depositing a high index cladding increases the bending losses and hence reduces the integration density. Finally, only positive or negative thermal shifts have been achieved thus far[26–28]; bidirectional tuning on a single platform remains elusive.

In this work we report for the first time that inductively-coupled plasma chemical vapor deposition (ICPECVD) can be used to tailor the thermo-optic properties of optical devices by depositing silicon dioxide claddings, the most common optical material, achieving large positive and negative thermal wavelength shifts on a single chip without significantly affecting the optical losses. We apply this technique on amorphous silicon carbide, silicon nitride and silicon-on-insulator platforms, and demonstrate an up to 10-fold improvement of the thermo-optical wavelength tunability of SiN compared to literature values. Thanks to the bidirectional tunability of our silicon dioxide, we demonstrate an almost 5-fold higher thermal tunability as well as athermal photonic ring resonators on a-SiC platform. This powerful tunability range allows us to showcase unprecedented photonic devices by deterministically including claddings with both negative and positive thermal response on the same chip. Moreover, thanks to our low temperature deposition technique, we introduce a novel fabrication approach to isolate active optical devices, and demonstrate a decrease in thermal crosstalk by at least two orders of magnitude.

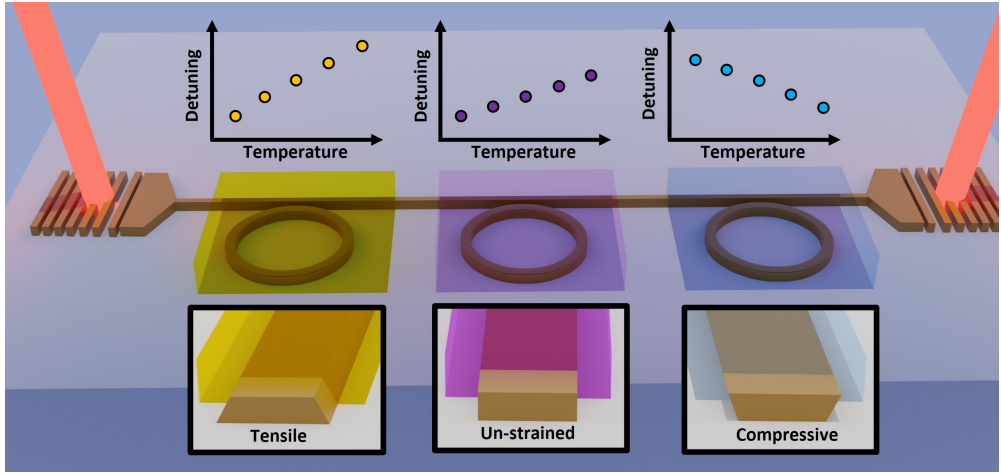

**Fig. 1 Illustration of different claddings deposited deterministically on a single chip resulting in different waveguide thermal strain and thermo-optic tunability.** Depositing SiO$_2$ claddings at different deposition temperatures results in tensile and compressive effects in the waveguides. This strain can be related to different behaviors of the detuning of the wavelength as the temperature is raised. An un-strained waveguide displays normal thermo-optic response. This is the case for e.g. PECVD-deposited SiO$_2$ cladding. The layer on top influences a thermal expansion mismatch between materials, changing the sign of the thermal shift. For devices made from a-SiC, the thermo-optic tunability can increase for tensile strain, while it becomes negative for compressive strain.

## 2 Results and discussion

### 2.1 Thermo-optic wavelength shift and propagation loss

As illustrated in Fig. 1 and demonstrated in other works, applying thermal stress in the cladding allows controlling the temperature sensitivity of optical waveguides[29]. In CVD techniques it is well known that different parameters such as temperature and chamber pressure can modify the stress profiles in the deposited films [30–32]. We measure the thermal response of optical ring resonators (radius 120 $\mu$m, waveguide width 750 nm, gap 850 nm and thickness 270 nm as measured by ellipsometry) fabricated on a-SiC films and covered with silicon dioxide claddings deposited via ICPCVD and PECVD techniques under different deposition temperatures (Fig. 2a) and chamber pressures (Fig. 2b). Details about free spectral range, group index, effective index, thermal shifts and device dimensions for a-SiC, SiN and Si platforms can be found in the Supplementary Information together with the calculated effective TOC. In Fig. 2a it can be seen that the largest thermal tunability of -166 pm/°C occurs for a deposition temperature of 75°C (final temperature of 91°C due to chamber heating). The spectra and fits can be found in the Supplementary Information. We find that the device response is not reproducible, resulting in different TOCs once metal heaters are placed or for different fabrication runs. To prevent this issue, the temperature to deposit these films should not be lower than the processing temperatures ($<$ 175°C). We therefore tune the thermal tunability at a fixed temperature by modifying the chamber pressure. At a deposition temperature of 150°C (Fig. 2b) we achieve shifts between +29.5 pm/°C (at 2 mTorr) and -118 pm/°C (at 16 mTorr). Using this approach, we record a thermal shift of -138 pm/°C in a-SiC platform depositing ICPCVD $SiO_2$ at 300°C and 12 mTorr, corresponding to $TOC_{eff}$ = -2.2×$10^{-4}$. The respective spectra at different temperatures can be seen in Fig. 2c. This represents a tunability of almost 5 times higher than standard devices [33] and significantly 22% more than that of silicon [34] (see Supplementary Information for the fitting). Crucially, for sensing applications, and thanks to the significant thermal tuning from negative to positive, our method allows for the fabrication of athermal devices by choosing the appropriate chamber pressure (3 mTorr) and deposition temperature (150°C). We achieved a thermal response as low as 1.5 pm/°C in a temperature range between 27°C and 35°C, a relevant temperature range for biological and chemical sensing (Fig. 2d), which is 20 times lower than the standard PECVD-cladded devices (see Supplementary Information for individual spectra).

To demonstrate that this method can be applied to other platforms, we deposit claddings using PECVD and ICPCVD at different temperatures on SOI and SiN platforms. The relative shifts as a function of temperature are shown in Fig. 2e and f, respectively. The representative spectra at different temperatures can be found in the Supplementary Information. Fig. 2e shows that for SOI optical ring resonators we can achieve thermal shifts between -96 pm/°C ($TOC_{eff}$ = -2.2×$10^{-4}$/°C) for ICPCVD oxide deposited at 75°C and +40 pm/°C for 300°C PECVD oxide cladding.

Similarly, Fig. 2f shows that depositing PECVD $SiO_2$ on SiN devices yields 14 pm/°C, comparable to values found in the literature[35]. In contrast, when this cladding is deposited with ICPCVD at 300°C we achieve a thermal shift of -106 pm/°C, representing an improvement of almost an order of magnitude and a $TOC_{eff}$ of -1.2×$10^{-4}$/°C.

In terms of optical quality, using ICPCVD $SiO_2$ at a temperature of 150°C and changing the chamber pressure results in devices with similar quality over the pressure range from 2 mTorr to 8 mTorr with waveguide propagation losses of 2.68 dB/cm ($Q_{int}$ = 1.58×$10^5$, comparable to literature values). At this deposition temperature, processing the devices at temperatures between 200°C to 400°C does not affect the thermo-optic properties of the devices (see Supplementary Information).

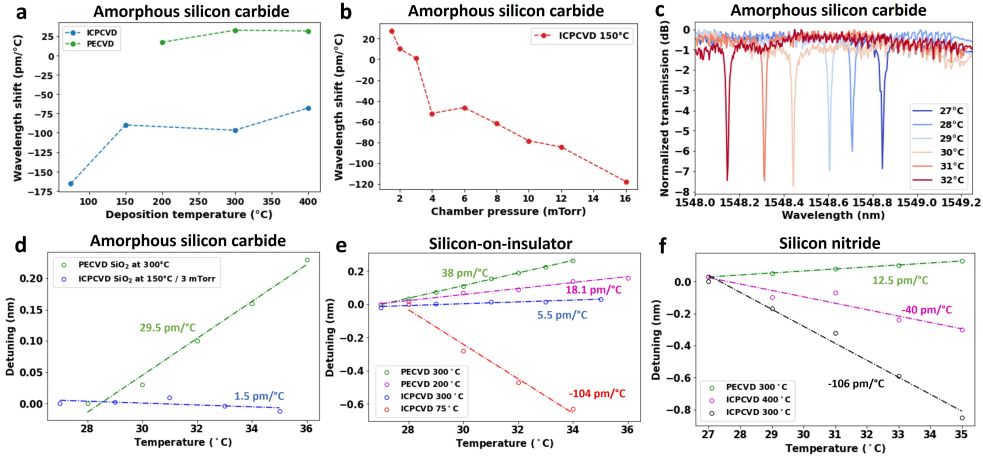

**Fig. 2 Thermo-optic tunability of devices made on a-SiC, SiN and SOI platforms using PECVD and ICPCVD claddings deposited at different conditions. a,** Wavelength shift in pm/°C for a-SiC devices with ICPCVD cladding deposited at different temperatures and a constant chamber pressure of 8 mTorr using ICPCVD (green) and PECVD (blue). Lines are guides to the eye. **b,** Wavelength shift in pm/°C for a-SiC devices with ICPCVD cladding deposited at different chamber pressures and a temperature of 150°C (red). **c,** Wavelength spectra of a device with a cladding deposited via ICPCVD at 300°C and 12 mTorr chamber pressure recorded at different temperatures between 27°C and 32°C. **d,** Relative wavelength shifts for a-SiC devices with cladding deposited by PECVD (chamber temperature 300°C; green symbols) and ICPCVD (chamber temperature 150°C and chamber pressure 3 mTorr; blue symbols). Lines are linear fits. **e,** Relative wavelength shifts of silicon-on-insulator optical devices with PECVD and ICPCVD claddings deposited at different temperatures. Lines are linear fits with slopes as indicated. **f,** Relative wavelength shifts of silicon nitride optical devices with PECVD and ICPCVD claddings deposited at different temperatures. Lines are linear fits with slopes as indicated.

## 2.2 Active and passive devices

To highlight the power of our fabrication strategy, we use it to create a coupled resonator optical waveguide (CROW) combining a positive and a negative thermal tunability without the need for separate heaters for each ring. This is done by depositing at the same temperature of 150°C, compatible with lithography resists, but with chamber pressures of 2 mTorr (positive TOC) and 8 mTorr (negative TOC) in accordance with Fig. 2b. CROW devices are typically used in optical filtering, dispersion compensation, and non-linear optics [36–39]. In addition, they can be used to delay, store, and buffer photons with controlled times[40]. The basic structure of a CROW device consists of two or more adjacent and coupled ring resonators.

In their usual configuration, these devices are tuned with separate micro-heaters on top of the rings. Once the coupling condition is fulfilled, the resonances of each ring overlap and the photons can be filtered to the output. In contrast, our CROW design incorporating two rings of opposite thermal shifts can be tuned by a single heater. Most importantly, the coupling condition can also be achieved without the use of any on-chip local heater which, to our knowledge, has not been demonstrated before. Using our fabrication scheme where the thermal tunability is controlled by the ICPCVD conditions, we fabricate a CROW configuration with two claddings (Fig. 3a). Fig. 3b shows an optical microscope image of the final rings with heaters. They are connected in series (or in parallel) having common voltage and common ground. For a device connected in parallel, sweeping the voltage results in the diagram shown in Fig. 3c. This figure represents a 2D map tracking the position of the dips for both rings as the voltage is increased. For a voltage of 6V, the resonance matching condition for both rings is fulfilled. The Supplementary Information includes a similar plot for a device connected in series, displaying similar behavior but requiring a higher operation voltage, as well as an example of a device with better thermal tunability. The differences in negative and positive wavelength shifts are not related to the material but to the misalignment of the micro-heaters and the waveguides. Passive actuation can also be performed by using a negative cladding in one of the devices and heating the entire sample stage. Fig. 3d shows a device with this configuration and Fig. 3e shows the output spectra as the temperature of the sample is increased from 32.5°C to 34.5°C. In this case, the matching condition is achieved at a temperature of 33.9°C. As another example, we show in the Supplementary Information a Mach-Zehnder interferometer where it is possible to vary the output power in the ports by modifying the stage temperature.

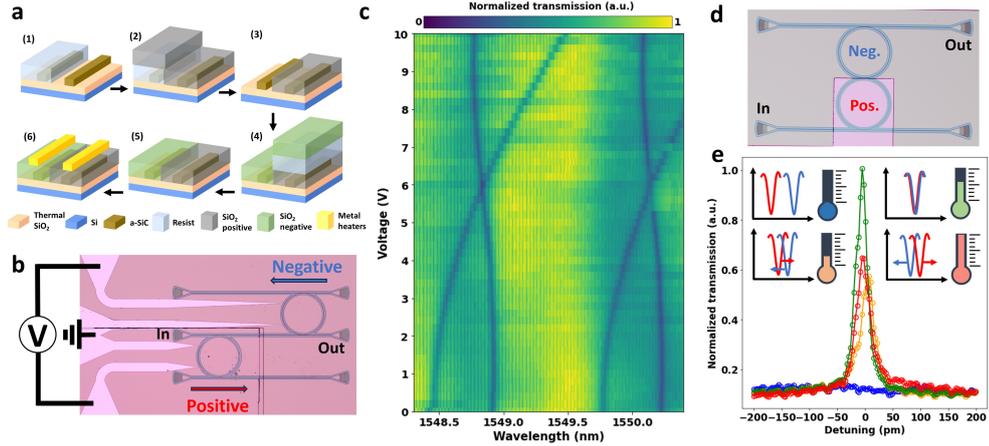

**Fig. 3 Fabrication route for the deterministic integration of silicon dioxide claddings with bidirectional thermal tunability and measurements of active and passive devices. a,** Fabrication scheme for the inclusion of bidirectional claddings in optical devices. 1) Resist spin coating, exposure and development. 2) SiO$_2$ cladding deposition for positive shift. 3) Lift-off in acetone. 4) Resist spin-coating, exposure, development and deposition of negative TOC cladding. 5) Lift-off in acetone. 6) Patterning of metal heaters via lift-off. **b,** Optical microscope image of two ring resonators connected with a middle waveguide. **c,** Resonant wavelength as a function of the voltage applied to metal heaters of the optical device shown in panel b. **d,** CROW resonator fabricated using bidirectional thermal response with only one cladding and **e,** Transmission spectra around 1552 nm of the output signal of the device shown in panel d at temperatures of 32.5°C (blue), 33.5°C (orange), 33.9°C (green) and 34.5°C (red).

### 2.3 Cladding lift-off

In standard methods, the cladding is deposited on the whole sample, thermally connecting the devices and making it impractical to place optical devices close to each other. Some approaches have shown that, to reduce the thermal crosstalk, the cladding between devices can be etched [41–43] or predictive models can be developed to control their overall response [44–46].

The low processing temperatures involved in ICPCVD allow us to perform lift-off to limit the cladding to a region of 4 $\mu$m around the waveguide. For reference, we also fabricate ring resonators with standard PECVD cladding and a 10 $\mu$m separation between the optical devices. Fig. 4a shows optical images of the devices together with a schematic side-view of the cladding. To determine the thermal crosstalk between adjacent rings, we vary the power dissipated in the heater of ring A and observe the thermal response of both rings A and B (Fig. 4a). In the case of standard PECVD cladding, the thermal response of rings A and B is 18.5 pm/mW (red line in Fig. 4b) and 1.5 pm/mW (red line in Fig. 4c), respectively. For ICPCVD cladding, the thermal response of rings A and B is -22.7 pm/mW (purple line in Fig. 4b) and -2.5 pm/mW (purple line in Fig. 4c), respectively. By using an ICPCVD cladding with lift-off method, we improve the performance of our device in two ways: we increase the thermal response in ring A (42 pm/mW, green line in Fig. 4b) and decrease the thermal response in ring B (no thermal shift, green line in Fig. 4c). Therefore, we can thermally isolate two ring resonators placed 10 $\mu$m apart by depositing an ICPCVD cladding with the lift-off method, which is not feasible in standard PECVD claddings (see the Supplementary Information for specific spectra). Note that the heating efficiency of the on-chip micro-heaters falls outside the scope of this study, and can be engineered to significantly raise the ratio between heat generation and power consumption, and optimize the reconfiguration time [47–51].

## 3 Conclusions

In conclusion, this is the first study and demonstration of ICPCVD silicon dioxide claddings deposited at low temperatures to achieve positive, negative and athermal thermo-optic devices on a single chip with large thermal tunability across several photonic platforms such as amorphous silicon carbide, silicon nitride and silicon-on-insulator. Most importantly, we fabricated both passive and active components. Our approach opens up low power photonic configurations such as Mach-Zhender interferometers, single-heater CROW optical devices, and highly sensitive temperature sensors that could be easily integrated with current electronic and photonic technologies. Additionally, we showed that the low-temperature fabrication scheme allows to thermally isolate the optical devices to compensate for the high thermal shifts and to increase the photonic integration densities. With further improvements, we foresee the use of these configurations for novel photonic architectures and highly tunable photonic circuits.

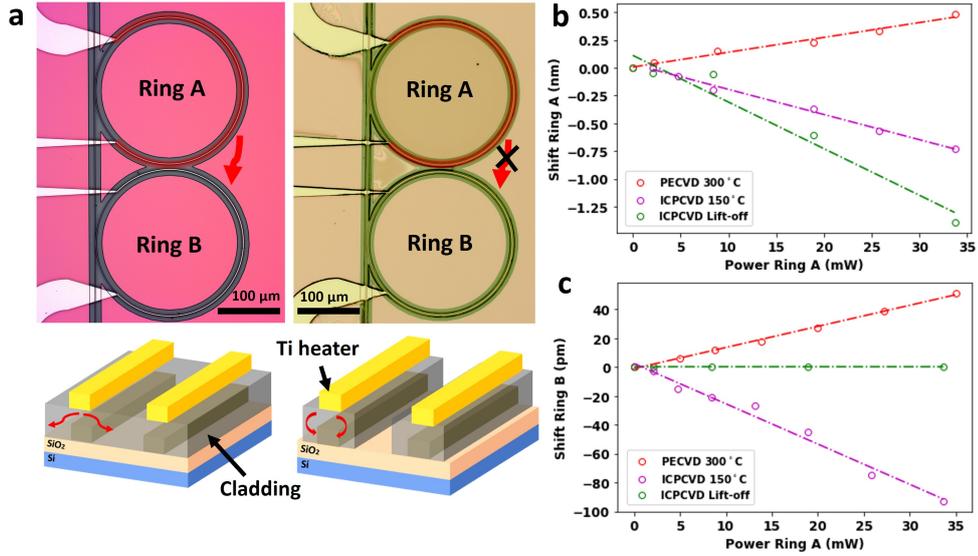

**Fig. 4 Thermal crosstalk measurements in ring resonators using claddings deposited by PECVD, ICPCVD and with ICPCVD lift-off. a,** Optical image (top view) of ring resonator devices fabricated using continuous cladding (left) and cladding delimited with lift-off (right). Red arrows denote thermal crosstalk. **b,** Detuning as a function of power consumed in ring A fabricated using PECVD (red), ICPCVD (purple) and cladding lift-off (green). **c,** Shift of the adjacent ring due to thermal crosstalk as a function of power consumed in the micro-heater for PECVD (red), ICPCVD (purple) and cladding lift-off (green).

# 4 Supplementary information

Detailed information about film characterization, optical setup, summarized data for each device, passive device configurations, strain release experiments, representative spectra for the different platforms, additional measurements for CROW devices, thermal crosstalk spectra and electron microscope image of the lifted-off cladding.

# 5 Acknowledgements

I.E.Z. acknowledges funding from the European Union's Horizon Europe research and innovation programme under grant agreement No. 101098717 (RESPITE project) and No. 101099291 (fastMOT project). I.E.Z and N.S. acknowledge the NWO OTP COMB-O project (18757). Z.L. acknowledges the China Scholarship Council (CSC, 202206460012).

# 6 Author contribution

I.E.Z, B.L.R., and N.S. conceived and coordinated the project. B.L.R, Z.L., and R.K. fabricated the samples. B.L.R. and N.S. characterized the samples. B.L.R. and Z.L. performed FDTD simulations. B.L.R. and R.K. performed ellipsometry, XRD, Raman spectroscopy, AFM and stress measurements of the films. J.B. and T.S. assisted in the adaptation of the optical setups. J.C. and S.G. provided SOI samples and assisted in the fabrication on this platform. All authors contributed to writing the article and reading and approving the final manuscript.

## Declarations

The authors declare no conflicts of interest.

## Appendix A  Deposition recipes for silicon dioxide

Using PECVD we deposited $SiO_2$ in a mixture of 8.5 sccm $SiH_4$, 710 sccm $N_2O$ and 165 sccm $N_2$ with a plasma power of 20 W and chamber pressure of 1000 mTorr. The average deposition rates for all recipes were around 70 nm/min. The deposition of $SiO_2$ claddings using ICPCVD was done using a mixture of 16 sccm $SiH_4$ and 60 sccm $N_2O$ with a plasma power of 1300W and a chamber pressure of 8 mTorr. The average deposition rates for all recipes were around 60 nm/min.

## Appendix B  Device fabrication

The fabrication of a-SiC optical devices was performed following our previous work [33]. The development of the electron beam resist was performed in pentyl-acetate, MBIK:IPA (1:1) and IPA. For the etching of a-SiC and silicon-on-insulator, the plasma power was set to 20W with 13.5 sccm of $SF_6$ and 3.5 sccm of $O_2$ at a chamber pressure of 8 $\mu$bar. LPCVD Silicon nitride devices were fabricated on 380 nm thick films. Etching of these devices was performed at 20°C in a mixture of $CHF_3$ and $O_2$. The excess e-beam resist was removed by oxygen plasma cleaning (200 sccm $O_2$ at 50W) for 5 minutes. The lift-off of silicon dioxide was performed using PMMA 950K A11 with a thickness of 2.2 $\mu$m. The resist was spin-coated at a rate of 2500 rpm/min and baked for 10 min at a temperature of 175°C. To remove resist residues after development, we performed an oxygen plasma cleaning for 60s (200 sccm $O_2$ at 50W). The lift-off was done by immersing the samples in acetone and heating the solution to 53°C. Silicon dioxide claddings were deposited via ICPCVD or PECVD with a thickness of 2 $\mu$m. The contact pads were made of 80 nm titanium and 10 nm gold and patterned using lift-off with PMMA A6-950K resist with a thickness of 1 $\mu$m.

## Appendix C  Device characterization

For the device characterization, we used a grating coupler configuration. The light from the laser source was coupled to a polarization-maintaining fiber using a U-Bench. The polarization going to the device was controlled with a free-space polarizer. Paddle polarizers (FPC560) were used to align the polarization of the laser source to the free-space polarizer (FBR-LPNIR). The thermal response of the devices was recorded using an optical spectrum analyzer (OSA Yokowaga AQ6374) and a tunable laser (Photonetics TUNICS-PRI 3642 HE 15) with a powermeter (818-IR Newport, linearity 0.5%). The sample was placed in a PCB using thermally conductive silver paste and heated with a thermal element in the sample stage with a temperature PIC controller (see Supplementary Information). Several graphs for the different platforms show the shift in resonance wavelength as a function of the stage temperature and are represented in the Supplementary Information. Electrical tuning of the devices was done with micro-heaters of 80 nm thick titanium (Ti) on top of the 2 $\mu$m SiO$_2$ cladding. The voltage was swept using a programmable voltage supply (RIGOL model DP832A) and a customized MATLAB script. The propagation losses were calculated using the intrinsic quality factor and group index of the devices following previous studies [33, 52].

# References

[1] Rizzo, A., Novick, A., Gopal, V., Kim, B.Y., Ji, X., Daudlin, S., Okawachi, Y., Cheng, Q., Lipson, M., Gaeta, A.L., *et al.*: Massively scalable kerr comb-driven silicon photonic link. Nature Photonics **17**(9), 781–790 (2023)

[2] Lipson, M.: The revolution of silicon photonics. Nature Materials **21**(9), 974–975 (2022)

[3] Miller, S.A., Yu, M., Ji, X., Griffith, A.G., Cardenas, J., Gaeta, A.L., Lipson, M.: Low-loss silicon platform for broadband mid-infrared photonics. Optica **4**(7), 707–712 (2017)

[4] Ji, X., Okawachi, Y., Gil-Molina, A., Corato-Zanarella, M., Roberts, S., Gaeta, A.L., Lipson, M.: Ultra-low-loss silicon nitride photonics based on deposited films compatible with foundries. Laser & Photonics Reviews **17**(3), 2200544 (2023)

[5] Ji, X., Roberts, S., Corato-Zanarella, M., Lipson, M.: Methods to achieve ultra-high quality factor silicon nitride resonators. APL Photonics **6**(7) (2021)

[6] Stern, B., Ji, X., Dutt, A., Lipson, M.: Compact narrow-linewidth integrated laser based on a low-loss silicon nitride ring resonator. Optics letters **42**(21), 4541–4544 (2017)

[7] Ji, X., Barbosa, F.A., Roberts, S.P., Dutt, A., Cardenas, J., Okawachi, Y., Bryant, A., Gaeta, A.L., Lipson, M.: Ultra-low-loss on-chip resonators with sub-milliwatt parametric oscillation threshold. Optica **4**(6), 619–624 (2017)

[8] Klenner, A., Mayer, A.S., Johnson, A.R., Luke, K., Lamont, M.R., Okawachi, Y., Lipson, M., Gaeta, A.L., Keller, U.: Gigahertz frequency comb offset stabilization based on supercontinuum generation in silicon nitride waveguides. Optics express **24**(10), 11043–11053 (2016)

[9] Luke, K., Okawachi, Y., Lamont, M.R., Gaeta, A.L., Lipson, M.: Broadband mid-infrared frequency comb generation in a si 3 n 4 microresonator. Optics letters **40**(21), 4823–4826 (2015)

[10] Yi, A., Wang, C., Zhou, L., Zhu, Y., Zhang, S., You, T., Zhang, J., Ou, X.: Silicon carbide for integrated photonics. Applied Physics Reviews **9**(3) (2022)

[11] Yi, A., Zheng, Y., Huang, H., Lin, J., Yan, Y., You, T., Huang, K., Zhang, S., Shen, C., Zhou, M., *et al.*: Wafer-scale 4h-silicon carbide-on-insulator (4h–sicoi) platform for nonlinear integrated optical devices. Optical Materials **107**, 109990 (2020)

[12] Zheng, Y., Pu, M., Yi, A., Ou, X., Ou, H.: 4h-sic microring resonators for nonlinear integrated photonics. Optics letters **44**(23), 5784–5787 (2019)

[13] Wang, C., Fang, Z., Yi, A., Yang, B., Wang, Z., Zhou, L., Shen, C., Zhu, Y., Zhou, Y., Bao, R., *et al.*: High-q microresonators on 4h-silicon-carbide-on-insulator platform for nonlinear photonics. Light: Science & Applications **10**(1), 139 (2021)

[14] Lukin, D.M., Dory, C., Guidry, M.A., Yang, K.Y., Mishra, S.D., Trivedi, R., Radulaski, M., Sun, S., Vercruysse, D., Ahn, G.H., Vučković, J.: 4h-silicon-carbide-on-insulator for integrated quantum and nonlinear photonics. Nature Photonics **14**(5), 330–334 (2019)

[15] Rabiei, P., Ma, J., Khan, S., Chiles, J., Fathpour, S.: Heterogeneous lithium niobate photonics on silicon substrates. Optics express **21**(21), 25573–25581 (2013)

[16] Pohl, D., Escalé, M.R., Madi, M., Kaufmann, F., Brotzer, P., Sergeyev, A., Guldimann, B., Giaccari, P., Alberti, E., Meier, U., Grange, R.: An integrated broadband spectrometer on thin-film lithium niobate. Nature Photonics **14**(1), 24–29 (2019)

[17] Feng, H., Ge, T., Guo, X., Wang, B., Zhang, Y., Chen, Z., Zhu, S., Zhang, K., Sun, W., Huang, C., et al.: Integrated lithium niobate microwave photonic processing engine. Nature, 1–8 (2024)

[18] Qiu, F., Spring, A.M., Yokoyama, S.: Athermal and high-q hybrid tio2–si3n4 ring resonator via an etching-free fabrication technique. ACS Photonics **2**(3), 405–409 (2015)

[19] Robinson, J.T., Chen, L., Lipson, M.: On-chip gas detection in silicon optical microcavities. Opt. Express **16**(6), 4296–4301 (2008)

[20] Robinson, J.T., Preston, K., Painter, O., Lipson, M.: First-principle derivation of gain in high-index-contrast waveguides. Opt. Express **16**(21), 16659–16669 (2008)

[21] Arbabi, A., Goddard, L.L.: Measurements of the refractive indices and thermo-optic coefficients of si3n4 and siox using microring resonances. Opt. Lett. **38**(19), 3878–3881 (2013)

[22] Elshaari, A.W., Zadeh, I.E., Jöns, K.D., Zwiller, V.: Thermo-optic characterization of silicon nitride resonators for cryogenic photonic circuits. IEEE Photonics Journal **8**(3), 1–9 (2016)

[23] Chang, L.-Y.S., Pappert, S., Yu, P.K.L.: High thermo-optic tunability in pecvd silicon-rich amorphous silicon carbide. Opt. Lett. **48**(5), 1188–1191 (2023)

[24] Nejadriahi, H., Friedman, A., Sharma, R., Pappert, S., Fainman, Y., Yu, P.: Thermo-optic properties of silicon-rich silicon nitride for on-chip applications. Opt. Express **28**(17), 24951–24960 (2020)

[25] Memon, F.A., Morichetti, F., Melloni, A.: High thermo-optic coefficient of silicon oxycarbide photonic waveguides. ACS Photonics **5**(7), 2755–2759 (2018)

[26] Park, H., Jung, J., Zhang, Y., Liu, M., Lee, J., Noh, H., Choi, M., Lee, S., Park, H.: Effects of thermally induced phase transition on the negative thermo-optic properties of atomic-layer-deposited tio2 films. ACS Applied Electronic Materials **4**(2), 651–662 (2021)

[27] Teng, J., Dumon, P., Bogaerts, W., Zhang, H., Jian, X., Han, X., Zhao, M., Morthier, G., Baets, R.: Athermal silicon-on-insulator ring resonators by overlaying a polymer cladding on narrowed waveguides. Optics express **17**(17), 14627–14633 (2009)

[28] Alipour, P., Hosseini, E.S., Eftekhar, A.A., Momeni, B., Adibi, A.: Temperature-insensitive silicon microdisk resonators using polymeric cladding layers. In: 2009 Conference on Lasers and Electro-Optics and 2009 Conference on Quantum Electronics and Laser Science Conference, pp. 1–2 (2009). IEEE

[29] Huang, M., Yan, X.: Thermal-stress effects on the temperature sensitivity of optical waveguides. J. Opt. Soc. Am. B **20**(6), 1326–1333 (2003)

[30] Guan, D., Bruccoleri, A., Heilmann, R., Schattenburg, M.: Stress control of plasma enhanced chemical vapor deposited silicon oxide film from tetraethoxysilane. Journal of Micromechanics and Microengineering **24**(2), 027001 (2013)

[31] Wei, J., Ong, P.L., Tay, F.E., Iliescu, C.: A new fabrication method of low stress pecvd sinx layers for biomedical applications. Thin Solid Films **516**(16), 5181–5188 (2008)

[32] Greenhorn, S., Bano, E., Stambouli, V., Zekentes, K.: Amorphous sic thin films deposited by plasma-enhanced chemical vapor deposition for passivation in biomedical devices. Materials **17**(5), 1135 (2024)

[33] Lopez-Rodriguez, B., Kolk, R., Aggarwal, S., Sharma, N., Li, Z., Plaats, D., Scholte, T., Chang, J., Gröblacher, S., Pereira, S.F., Bhaskaran, H., Zadeh, I.E.: High-quality amorphous silicon carbide for hybrid photonic integration deposited

at a low temperature. ACS Photonics **10**(10), 3748–3754 (2023)

[34] Pruessner, M.W., Stievater, T.H., Ferraro, M.S., Rabinovich, W.S.: Thermo-optic tuning and switching in soi waveguide fabry-perot microcavities. Optics Express **15**(12), 7557–7563 (2007)

[35] Ilie, S.T., Faneca, J., Zeimpekis, I., Bucio, T.D., Grabska, K., Hewak, D.W., Chong, H.M., Gardes, F.Y.: Thermo-optic tuning of silicon nitride microring resonators with low loss non-volatile sb 2 s 3 phase change material. Scientific Reports **12**(1), 17815 (2022)

[36] Kumar, R.R., Tsang, H.K.: High-extinction crow filters for scalable quantum photonics. Optics Letters **46**(1), 134–137 (2021)

[37] Hryniewicz, J., Absil, P., Little, B., Wilson, R., Ho, P.-T.: Higher order filter response in coupled microring resonators. IEEE Photonics Technology Letters **12**(3), 320–322 (2000)

[38] Little, B., Chu, S., Pan, W., Ripin, D., Kaneko, T., Kokubun, Y., Ippen, E.: Vertically coupled glass microring resonator channel dropping filters. IEEE Photonics Technology Letters **11**(2), 215–217 (1999)

[39] Madsen, C., Lenz, G.: Optical all-pass filters for phase response design with applications for dispersion compensation. IEEE Photonics Technology Letters **10**(7), 994–996 (1998)

[40] Takesue, H., Matsuda, N., Kuramochi, E., Munro, W.J., Notomi, M.: An on-chip coupled resonator optical waveguide single-photon buffer. Nature communications **4**(1), 2725 (2013)

[41] Gilardi, G., Yao, W., Haghighi, H.R., Leijtens, X.J., Smit, M.K., Wale, M.: Deep trenches for thermal crosstalk reduction in inp-based photonic integrated circuits. Journal of lightwave technology **32**(24), 4864–4870 (2014)

[42] Ceccarelli, F., Atzeni, S., Pentangelo, C., Pellegatta, F., Crespi, A., Osellame, R.: Low power reconfigurability and reduced crosstalk in integrated photonic circuits fabricated by femtosecond laser micromachining. Laser & Photonics Reviews **14**(10), 2000024 (2020)

[43] Wu, Q., Zhou, L., Sun, X., Zhu, H., Lu, L., Chen, J.: Silicon thermo-optic variable optical attenuators based on mach–zehnder interference structures. Optics Communications **341**, 69–73 (2015)

[44] Milanizadeh, M., Aguiar, D., Melloni, A., Morichetti, F.: Canceling thermal crosstalk effects in photonic integrated circuits. Journal of Lightwave Technology **37**(4), 1325–1332 (2019)

[45] Teofilovic, I., Cem, A., Sanchez-Jacome, D., Perez-Lopez, D., Da Ros, F.: Thermal crosstalk modelling and compensation methods for programmable photonic integrated circuits. arXiv preprint arXiv:2404.10589 (2024)

[46] Orlandin, M., Cem, A., Curri, V., Carena, A., Da Ros, F., Bardella, P.: Thermal crosstalk effects in a silicon photonics neuromorphic network. In: 2023 International Conference on Numerical Simulation of Optoelectronic Devices (NUSOD), pp. 43–44 (2023). IEEE

[47] Atabaki, A., Hosseini, E.S., Eftekhar, A., Yegnanarayanan, S., Adibi, A.: Optimization of metallic microheaters for high-speed reconfigurable silicon photonics. Optics express **18**(17), 18312–18323 (2010)

[48] Cao, L., Aboketaf, A.A., Preble, S.F.: Cmos compatible micro-oven heater for efficient thermal control of silicon photonic devices. Optics Communications **305**, 66–70 (2013)

[49] Alemany, R., Muñoz, P., Pastor, D., Domínguez, C.: Thermo-optic phase tuners analysis and design for process modules on a silicon nitride platform. In: Photonics, vol. 8, p. 496 (2021). MDPI

[50] Erickson, J.R., Shah, V., Wan, Q., Youngblood, N., Xiong, F.: Designing fast and efficient electrically driven phase change photonics using foundry compatible waveguide-integrated microheaters. Optics Express **30**(8), 13673–13689 (2022)

[51] Li, Z., Chen, H., Wang, J., Lu, H., Liu, C.: Compact design of an optical phase shifter packaged with ist microheater used for integrated photonics. Results in Physics **19**, 103644 (2020)

[52] Barclay, P.E., Srinivasan, K., Painter, O.: Nonlinear response of silicon photonic crystal microresonators excited via an integrated waveguide and fiber taper. Optics express **13**(3), 801–820 (2005)

# Supplementary information: Magic silicon dioxide for widely tunable photonic integrated circuits

Bruno Lopez-Rodriguez[1], Naresh Sharma[1], Zizheng Li[1], Roald van der Kolk[1], Jasper Van Der Boom[1], Thomas Scholte[1], Jin Chang[2], Simon Gröblacher[2] and Iman Esmaeil Zadeh[1]

[1]Department of Imaging Physics (ImPhys), Faculty of Applied Sciences, Delft University of Technology, Delft 2628 CJ, The Netherlands
[2]Department of Quantum Nanoscience, Faculty of Applied Sciences, Delft University of Technology, Delft 2628 CJ, The Netherlands

E-mail: b.lopezrodriguez@tudelft.nl

**Table of Contents**

1. Film characterization

This section contains stress measurements, atomic force microscope and ellipsometry data of the deposited films of silicon dioxide deposited with ICPCVD and PECVD techniques at different temperatures.

1.1. Film stress

We measured the stress of amorphous silicon carbide and silicon dioxide films using FLX-2320-S Thin Film Stress Measurement system from Toho Technology. Silicon carbide films were deposited on top of thermally oxidized silicon wafers (525 µm) with oxide thickness of 8 µm. To characterize the silicon dioxide films they were deposited on bare silicon wafers. The data is summarized in **table S1**.

| Technique and condition | Stress (MPa) |
|---|---|
| a-SiC PECVD 300°C | -50 |
| a-SiC ICP 150°C | -300 to -500 |
| $SiO_2$ ICP 30°C | 10 |
| $SiO_2$ ICP 150°C | 10.4 |
| $SiO_2$ ICP 300°C | -22.9 |
| $SiO_2$ PECVD 300°C | -20 |

**Table S1.** Stress data for a-SiC and $SiO_2$ deposited using PECVD and ICPCVD techniques.

1.2. Atomic Force Microscopy

We performed AFM scans of the deposited Silicon Dioxide films with both techniques and retrieved surface morphology data, mainly surface roughness and skewness. The latter measures whether the surface has more deep valleys (negative skew) or protruding narrow peaks (positive). Three examples of AFM scans taken at different temperatures and techniques can be seen in **fig.S1**.

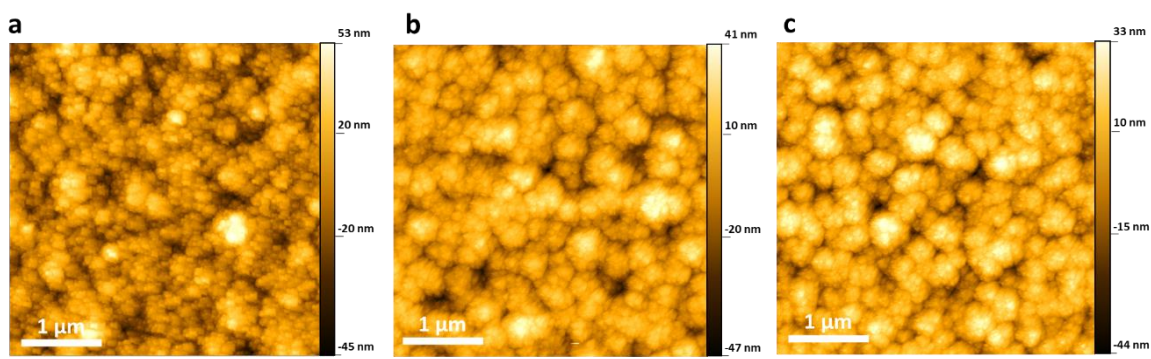

**Figure S1.** Atomic Force Microscope scans of silicon oxide deposited a) with PECVD at 300°C, b) with ICPCVD at 30°C and c) with ICPCVD at 150°C.

1.3. Ellipsometry

To characterize the refractive index of the silicon dioxide films deposited via ICPCVD and PECVD, we used Woollam M-2000 spectroscopic ellipsometer and fitted the corresponding data with a Cauchy model. **Table S2** summarizes the thickness, refractive index, surface roughness and skew obtained for the different films.

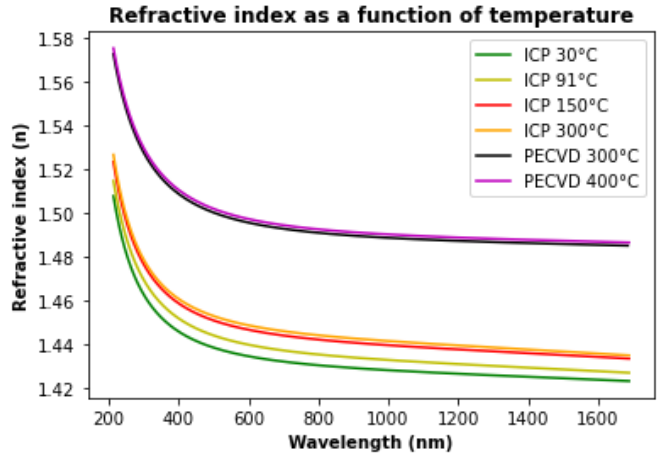

**Figure S2.** Ellipsometry data for the refractive index as a function of wavelength for PECVD and ICPCVD Silicon Dioxide films deposited at different temperatures.

| Parameter / Technique | Thickness (nm) | Refractive index | Surface roughness (RMS nm) | Skew |
|---|---|---|---|---|
| ICP 30°C | 2786.58 | 1.433 | 10.19 ± 1.34 | -0.99 |
| ICP 91°C | 2746.93 | 1.443 | --- | --- |
| ICP 150°C | 2641.77 | 1.445 | 9.55 ± 1.28 | -0.09 |
| ICP 300°C | 2607.79 | 1.447 | 11.49 ± 1.66 | -0.12 |
| PECVD 300°C | 3101.33 | 1.494 | 12.11 ± 1.91 | 0.19 |
| PECVD 400°C | 3284.44 | 1.496 | 10.75 ± 1.49 | 0.01 |

**Table S2.** Data for silicon dioxides deposited with ICPCVD and PECVD at different temperatures representing film thickness, refractive index, surface roughness and AFM skew.

## 2. Characterization setup and summarized data
### 2.1. Characterization setup

An schematic of the characterization setup described in the main manuscript is shown in **fig. S3a** with a picture of the optical setup in the lab shown in **fig.S3b**.

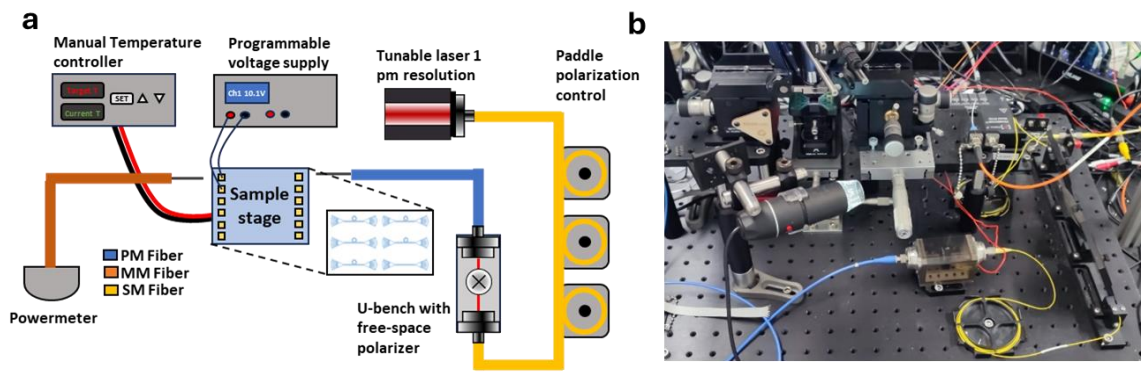

**Figure S3. a)** Schematic of the optical setup used for the measurements and **b)** full optical setup in the lab.

## 2.2. Optical properties of fabricated devices

For the silicon dioxide claddings deposited at different temperatures via ICPCVD and PECVD on a-SiC, SiN and SOI, we summarize in **table S3** the refractive index, free spectral range and corresponding group index of optical devices with dimensions stated in **table S5**.

|  | SiO$_2$ T (°C) | n oxide | a-SiC | | SiN | | SOI | |
|---|---|---|---|---|---|---|---|---|
|  |  |  | FSR (nm) | n$_g$ | FSR (nm) | n$_g$ | FSR (nm) | n$_g$ |
| **PECVD** | 200 | --- | 1.32 | 2.41 | --- | --- | 0.96 | 3.32 |
|  | 300 | 1.49440 | 1.30 | 2.45 | 1.49-1.55 | 2.14* | 0.96 | 3.32 |
|  | 400 | 1.49605 | 1.22 | 2.61 | --- | --- | --- |  |
| **ICP** | 30 | 1.43336 | 1.34 | 2.38 | 1.83 | 1.74 | 0.87 | 3.66 |
|  | 75 | --- | --- | --- | --- | --- | 0.90 | 3.54 |
|  | 150 | 1.44543 | 1.32 | 2.41 | 1.69-1.59 | 1.89* | 0.86 | 3.71 |
|  | 300 | 1.49605 | 1.30 | 2.45 | 1.80 | 1.68 | 0.86 | 3.71 |
|  | 400 | --- | 1.31 | 2.43 | 1.71-1.31 | 1.86* | 0.87 | 3.66 |

**Table S3.** Optical properties of the devices fabricated using different oxides with PECVD and ICPCVD at different temperatures. All the free spectral ranges are taken from the spectral measurement done at room temperature. *The corresponding group index is calculated assuming the FSR of the first mode.

For silicon dioxide claddings deposited via ICPCVD on a-SiC platform at a fixed deposition temperature of 150°C, we summarize in **table S4** the free spectral range and group index of the fabricated devices.

| Oxide pressure (mTorr) | 1.5 | 2 | 2.5 | 4 | 6 | 8 | 10 | 12 | 16 |
|---|---|---|---|---|---|---|---|---|---|
| FSR (nm) | --- | 1.32 | 1.34 | 1.35 | 1.36 | 1.32 | 1.43 | --- | 1.395 |
| n$_g$ |  | 2.41 | 2.38 | 2.36 | 2.34 | 2.41 | 2.23 | --- | 2.28 |

**Table S4.** Optical properties of the devices made with Silicon Dioxide deposited via ICPCVD at 150°C varying the chamber pressure.

To determine the effective thermo-optic coefficient, we simulated the mode profile with the selected dimensions using Ansys Lumerical MODE solutions for the different platforms. The table below summarizes information about width, thickness, ring radius, refractive index and obtained effective index. In **table S6** we summarize the mode overlap factor for the different platforms using the simulation results in **table S5**.

| Material | Width / Thickness (nm) | Ring radius (µm) | n | n$_{eff}$ |
|---|---|---|---|---|
| ICPCVD a-SiC 150°C | 800 / 271 | 120 | 2.67 | 1.937 |
| SOI | 700 / 220 | 120 | 3.44 | 2.565 |
| Silicon Nitride | 1000 / 368 | 120 | 2 | 2.014 |

**Table S5.** For the studied ring resonators summary of material platform, waveguide dimensions, ring radius and effective index calculated using FDTD (Ansys Lumerical MODE solutions).

| Material | Overlap waveguide (%) | Overlap cladding (%) | Overlap substrate (%) |
|---|---|---|---|
| ICPCVD a-SiC 150°C | 71.4856 | 13.7907 | 14.7237 |
| SOI | 79.4482 | 9.644 | 10.9078 |
| Silicon Nitride | 82.3079 | 8.73259 | 8.95951 |

**Table S6.** For the studied ring resonators mode fill factors calculated using FDTD (Ansys Lumerical MODE solutions).

**Table S7** summarizes the wavelength shifts in pm/°C and the corresponding effective thermos-optic coefficient for the different claddings deposited on a-SiC, SiN and SOI.

| SiO$_2$ T (°C) | | a-SiC | | SiN | | SOI | |
| --- | --- | --- | --- | --- | --- | --- | --- |
| Technique | | Shift (pm/°C) | TOC$_{eff}$ (10$^{-5}$) | Shift (pm/°C) | TOC$_{eff}$ (10$^{-5}$) | Shift (pm/°C) | TOC$_{eff}$ (10$^{-5}$) |
| PECVD | 200 | 16.87 | 2.12 | --- | --- | 17.1 | 3.00 |
| | 300 | 32.44 | 4.63 | 14 | 1.41 | 38 | 7.47 |
| | 400 | 31.37 | 4.78 | --- | --- | --- | --- |
| ICPCVD | 30 | 203 | 30.64 | -157 | -18.16 | 26.9 | 5.69 |
| | 75 | --- | --- | --- | --- | -94.6 | -22.28 |
| | 150 | -90 | -14.52 | -51 | -6.73 | 7.9 | 1.22 |
| | 300 | -96.7 | -15.80 | -86 | -9.83 | 4.8 | 0.48 |
| | 400 | -68.12 | -11.19 | -37 | -4.97 | 17.4 | 3.45 |

**Table S7.** For a-SiC, SiN and SOI optical ring resonators, thermal shift in pm/°C and effective thermo-optic coefficient for different oxide temperatures and techniques.

For the ICPCVD silicon dioxide claddings deposited at a temperature of 150°C and varying the chamber pressure, we summarize in **table S8** the wavelength shift in pm/°C and the corresponding effective thermos-optic coefficient.

| Oxide pressure (mTorr) | 1.5 | 2 | 2.5 | 4 | 6 | 8 | 10 | 12 | 16 |
| --- | --- | --- | --- | --- | --- | --- | --- | --- | --- |
| Shift (pm/°C) | 27.5 | 10.2 | 21.6 | -52.1 | -46.5 | -62 | -78.5 | -84.5 | -117.7 |
| TOC$_{eff}$ (10$^{-5}$) | --- | 1.09 | 2.81 | -8.44 | -7.53 | -10.02 | -11.18 | -16.74 | -17.85 |

**Table S8.** For a-SiC optical ring resonators, thermal shift in pm/°C and effective thermo-optic coefficient with an ICPCVD oxide cladding deposited at 150°C at different chamber pressures.

To determine the losses introduced by the different claddings deposited at different deposition temperatures and pressures on a-SiC optical devices, we summarized the main parameters of the analysed resonance dip together with the optical losses in dB/cm in **table S9**.

| Temperature (°C) | 30 | 150 | 300 |
| --- | --- | --- | --- |
| Full-Width at Half-Maximum (FWHM - pm) | 14.5 | 11.88 | 13.77 |
| Wavelength (nm) | 1549.73 | 1548.50 | 1549.87 |
| Transmission (au) | 0.367 | 0.412 | 0.46 |
| Loaded quality factor | 107,000 | 130,000 | 113,000 |
| Intrinsic quality factor | 133,000 | 158,000 | 134,000 |
| Group index | 2.378 | 2.41 | 2.45 |
| Loss (dB/cm) | 3.15 | 2.68 | 3.22 |

**Table S9.** Optical data for different ICPCVD Silicon Dioxide at a chamber pressure of 8 mTorr and different deposition temperatures of specific transmission dips.

We also deposited one of the ICPCVD claddings at a temperature of 300°C with a chamber pressure of 12 mTorr on a-SiC, corresponding to the wavelength spectra reported in the main manuscript (**fig.2c**). The data to determine the losses can be found in **table S10** has been taken from the spectra at room temperature (27°C).

| T (°C) | P (mTorr) | FWHM (pm) | Wavelength (nm) | T (au) | Q$_{load}$ | Q$_{int}$ | n$_g$ | Loss (dB/cm) |
| --- | --- | --- | --- | --- | --- | --- | --- | --- |
| 300 | 12 | 22.45 | 1550.22 | 0.265 | 69,000 | 91,000 | 2.41 | 4.74 |

**Table S10.** Optical data for ICPCVD Silicon Dioxide deposited at a temperature of 300°C and chamber pressure of 12 mTorr of a specific transmission dip.

**Table S11** summarizes main parameters of the measured resonance for a-SiC devices depositing silicon dioxide claddings via ICPCVD at a deposition temperature of 150°C and varying chamber pressure from 2.5 mTorr to 16 mTorr.

| Pressure (mTorr) | 2.5 | 8 | 10 | 16 |
|---|---|---|---|---|
| Full-Width at Half-Maximum (FWHM) | 13.07 | 11.88 | 18.80 | 30.27 |
| Wavelength (nm) | 1548.32 | 1548.50 | 1548.38 | 1550.4 |
| Transmission (au) | 0.329 | 0.412 | 0.363 | 0.250 |
| Loaded quality factor | 118,000 | 130,000 | 82,360 | 51,220 |
| Intrinsic quality factor | 150,000 | 158,000 | 103,000 | 68,000 |
| Group index | 2.38 | 2.41 | 2.23 | 2.28 |
| Loss (dB/cm) | 2.79 | 2.69 | 3.81 | 5.90 |

**Table S11.** Optical data for ICPCVD Silicon Dioxide deposited at a temperature of 150°C and different chamber pressures of a specific transmission dip.

**Fig. S4** shows a graphical representation of the change of free-spectral range, quality factor and refractive index for ICPCVD films deposited at different temperatures and pressures. **Fig.S4c** also shows the refractive index of silicon dioxide claddings deposited via PECVD at 300°C and 400°C.

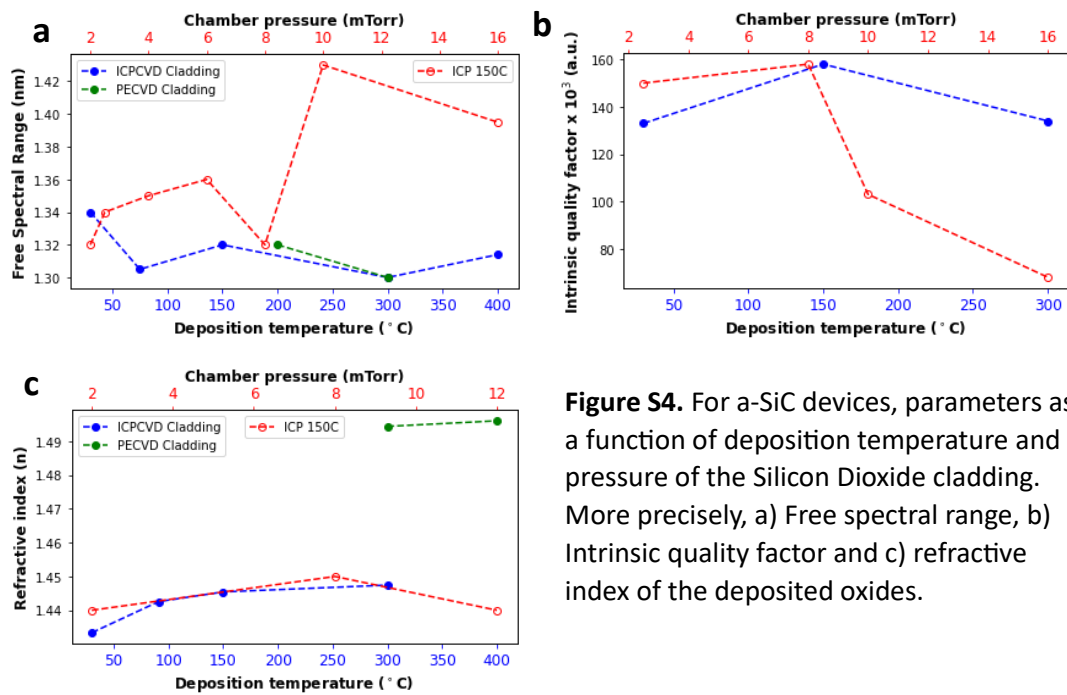

**Figure S4.** For a-SiC devices, parameters as a function of deposition temperature and pressure of the Silicon Dioxide cladding. More precisely, a) Free spectral range, b) Intrinsic quality factor and c) refractive index of the deposited oxides.

## 3. Passive devices

We demonstrate two different passive configurations of optical devices. **Fig.S5a** shows the resulting spectra two ring resonators connected in series with positive and negative claddings as the temperature of the sample stage is raise from 20°C to 37°C in steps of 2°C. **Fig.S5b** shows an optical microscope image of a Mach-Zehnder interferometer where one of the arms is covered with a cladding deposited via ICPCVD at a temperature of 150°C and chamber pressure of 8 mTorr. Raising the stage temperature vary the relative phase between each arm and the intensities at different outputs.

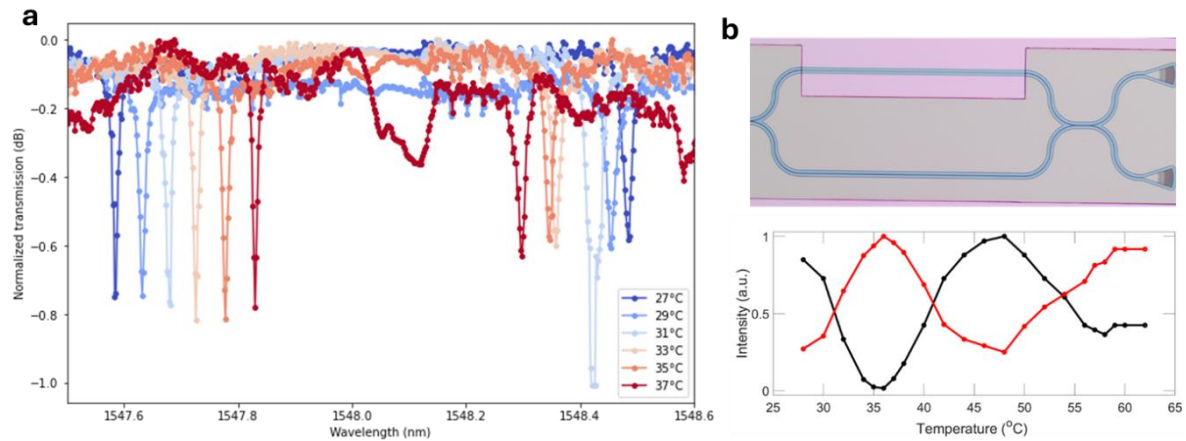

**Figure S5. a,** Spectra of ring resonators at different temperatures fabricated on the same chip with shared bus waveguide and claddings presenting bidirectional thermal response and **b,** Optical microscope image of a Mach-Zehnder interferometer covered with ICPCVD $SiO_2$ cladding in one of the arms and intensity as a function of temperature for the two output ports.

## 4. Strain release
### 4.1. Deposition of PECVD films

As an experiment to determine the effect in the thermal tunability when depositing other films with opposite thermal expansion, we deposited PECVD claddings on top of the ICPCVD cladded devices deposited at 150°C. The resonance wavelength position as a function of the stage temperature for the different configurations is depicted in **fig.S6**. It is observed that the dominant shift is similar to the one introduced by only using PECVD cladding and it cannot be attributed to annealing effects in the films as shown in section 4.3.

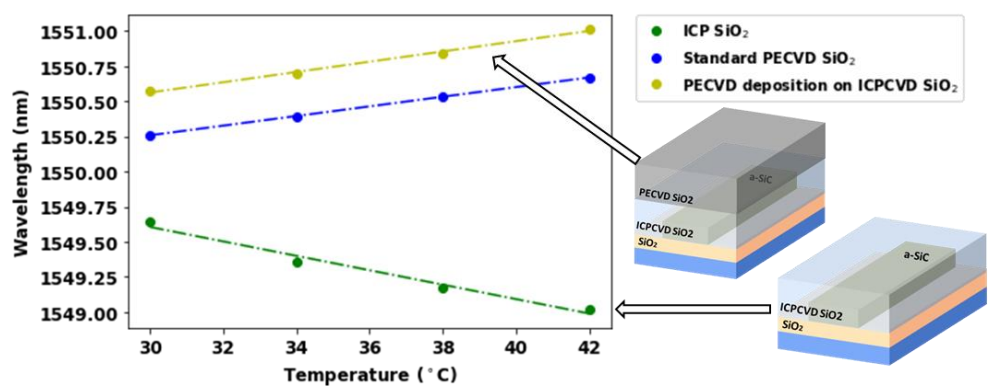

**Figure S6.** Wavelength shift as a function of temperature for optical ring resonators covered with ICPCVD $SiO_2$, Standard PECVD $SiO_2$ and PECVD on top of ICPCVD oxide.

## 4.2. Low temperature cladding

We deposited a silicon dioxide cladding via ICPCVD at 30°C on silicon nitride devices and the wavelength spectra as a function of the stage temperature between 27°C and 35°C is shown in **fig.S7**. When measuring the thermal response of the resonance, we observed that for stage temperatures higher than 33°C, there is a non-linear jump and the thermal tunability becomes lower. We attribute this effect to strain release between the core and the cladding that causes a decrease in the thermal tunability of the optical devices.

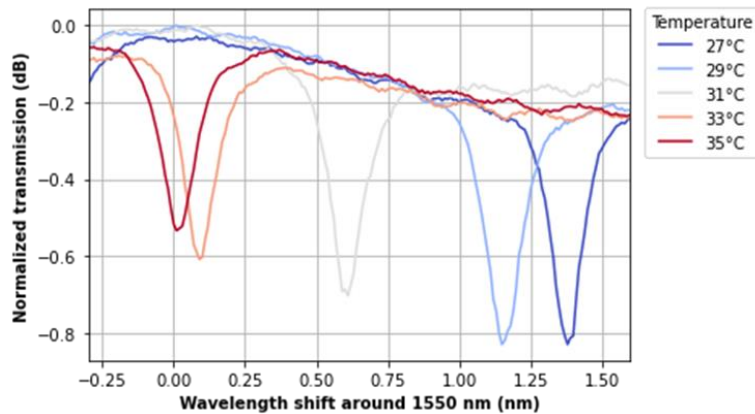

**Figure S7.** Wavelength spectra around 1550 nm for ICPCVD 30°C Silicon Dioxide cladding on a SiN device as a function of temperature between 27°C and 35°C in steps of 2°C.

## 4.3. Temperature stability

On films deposited via ICPCVD at a temperature of 150°C and chamber pressure of 8 mTorr, we also performed high temperature processing of the devices to investigate possible changes in the thermo-optic shift. The temperature range was done in incremental steps from 200°C to 400°C during 1h and the resulting position of the resonance dip as a function of stage temperature for the different processing temperatures is shown in **fig.S8**. We observed no difference in the thermal tunability.

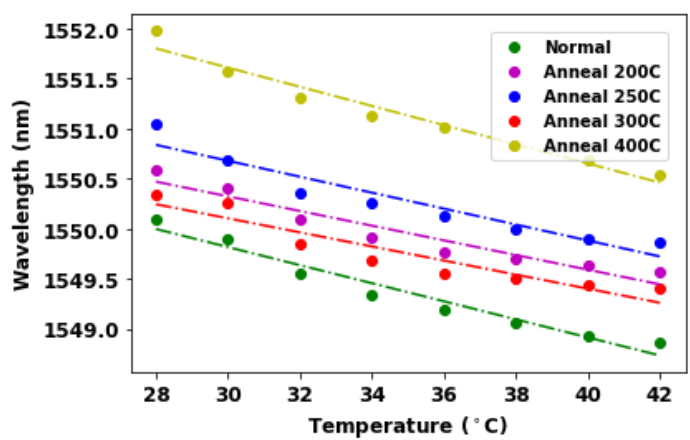

**Figure S8.** Wavelength shift as a function of the stage temperature of the selected resonance dip for different temperatures, from the standard devices (green) to annealing at 400°C (yellow).

## 5. Representative spectra for the different platforms

Below we include representative data and spectra for the wavelength shift of devices on amorphous silicon carbide, silicon nitride and silicon-on-insulator using different deposition conditions, mainly chamber pressure and deposition temperature.

### 5.1. Amorphous Silicon Carbide

In **fig.S9** we represent the wavelength shift as a function of stage temperature for ICPCVD silicon dioxide claddings deposited at 150°C and chamber pressures of 1.5 mTorr and 12 mTorr corresponding to negative and positive thermal tunability. **Fig.S10** shows the resulting wavelength spectra taken at different stage temperatures and linear fitting to obtain the wavelength tunability for a-SiC devices with silicon dioxide ICPCVD claddings deposited at 30°C and chamber pressure of 8 mTorr. This device presents the largest shift in wavelength in a stage temperature range between 27°C and 30°C.

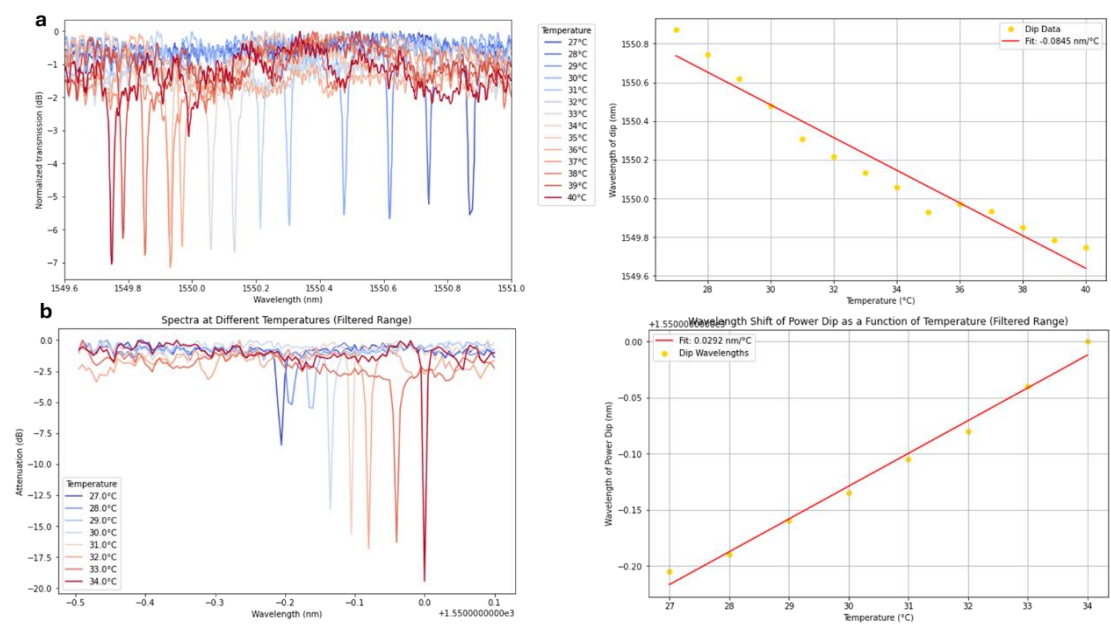

**Figure S9.** Spectra taken at different temperatures for a-SiC ring resonators with SiO$_2$ cladding deposited via ICPCVD at 150°C and corresponding fitting for chamber pressure of **a,** 12 mTorr and **b,** 1.5 mTorr.

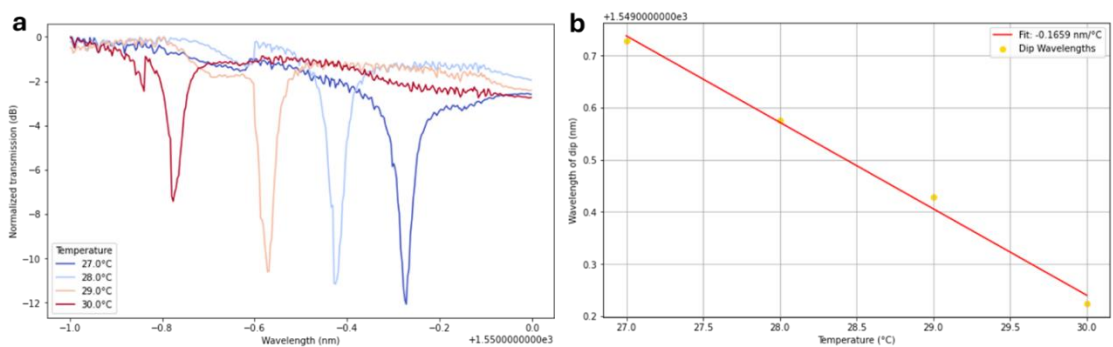

**Figure S10. a)** Normalized spectra at different stage temperatures for an a-SiC optical device with ICPCVD SiO$_2$ cladding deposited via ICPCVD at 75°C (ramp up to 91°C) and chamber pressure of 8 mTorr and **b)** resonance dip position with linear fitting.

We also include in **fig.S17** the wavelength spectra at different stage temperatures and corresponding resonance wavelength fitting of ICPCVD claddings deposited via ICPCVD at 150°C and varying the chamber pressure.

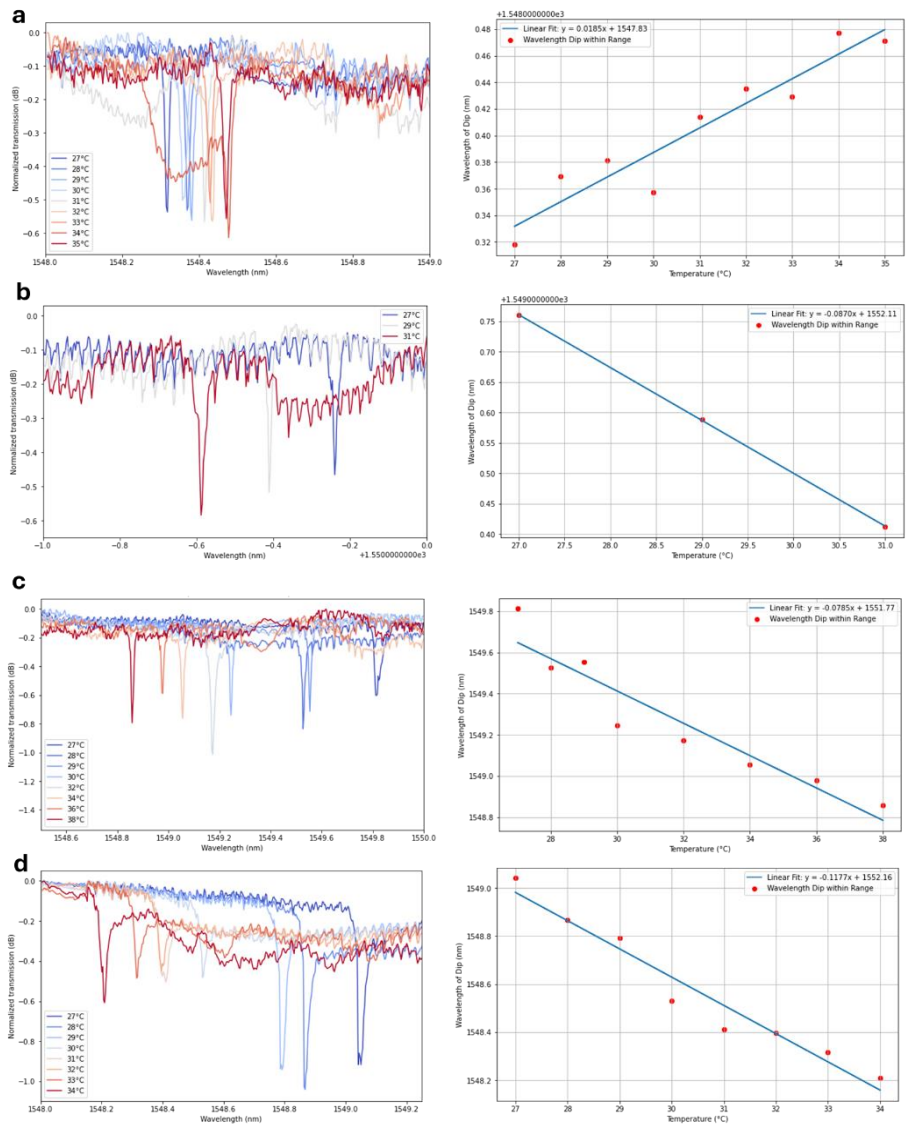

**Figure S11.** Spectra taken at different temperatures for a-SiC ring resonators with SiO$_2$ cladding deposited via ICPCVD at 150°C and corresponding fitting for chamber pressure of **a,** 2.5 mTorr and **b,** 8 mTorr, **c,** 10 mTorr and **d,** 16 mTorr.

The device that presented the highest negative shift for the a-SiC platform was fabricated using ICPCVD SiO$_2$ deposited at a temperature of 300°C and chamber pressure of 12 mTorr. The fitting of the resonance dip as a function of temperature is represented in **fig.S12** and corresponds to the data shown in **fig.2b** in the main manuscript.

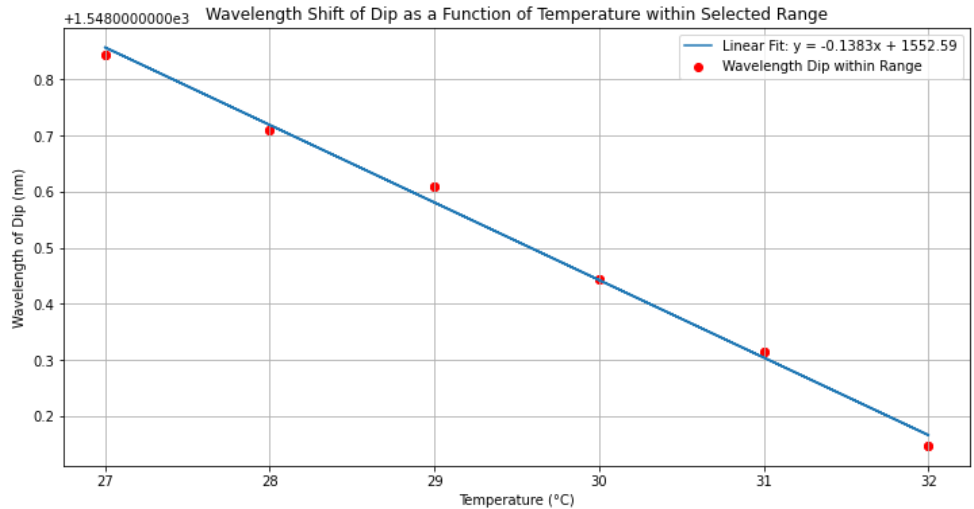

**Figure S12.** Wavelength of the selected resonance dip position as a function of temperature together with the linear fitting.

Attending to the systematic measurements done for ICPCVD claddings deposited at a temperature of 150°C and varying the chamber pressure on a-SiC optical devices, the athermal condition can be achieved for a chamber pressure of 3 mTorr. We deposited a silicon dioxide cladding on a-SiC with these conditions and measured the ring resonator in a temperature range between 27°C and 41°C in steps of 1°C. The resulting spectra is shown in **fig.13a** with the corresponding wavelength position as a function of the stage temperature in **fig.13b**. The thermal tunability between 27°C and 35°C obtained from fitting the data is 1.1 pm/°C . In the main article, the same data is represented in steps of 2°C resulting in a thermal tunability of 1.5 pm/°C.

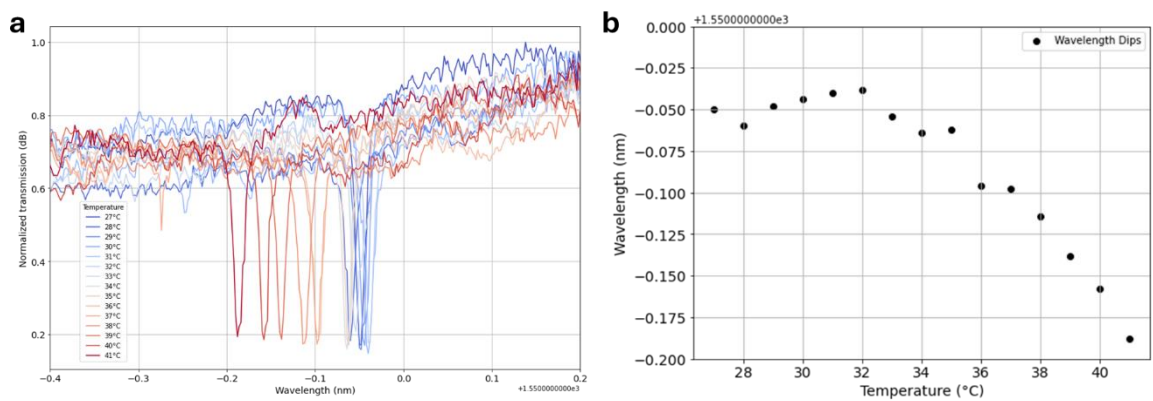

**Figure S13.** For an a-SiC ring resonator with ICPCVD SiO$_2$ cladding deposited at 150°C and chamber pressure of 3 mTorr shown in figure 2 of the manuscript **a,** Normalized spectra taken at different temperatures and **b,** dip position as a function of temperature.

## 5.2. Silicon Nitride devices

We deposited silicon dioxide claddings on silicon nitride devices via ICPCVD and PECVD at different temperatures and the resulting wavelength spectra as a function of stage temperature is found in **fig.S14.** As a reference, we also included the effect of a common electron beam resist PMMA on the thermal shift, known to give negative thermo-optic tunability.

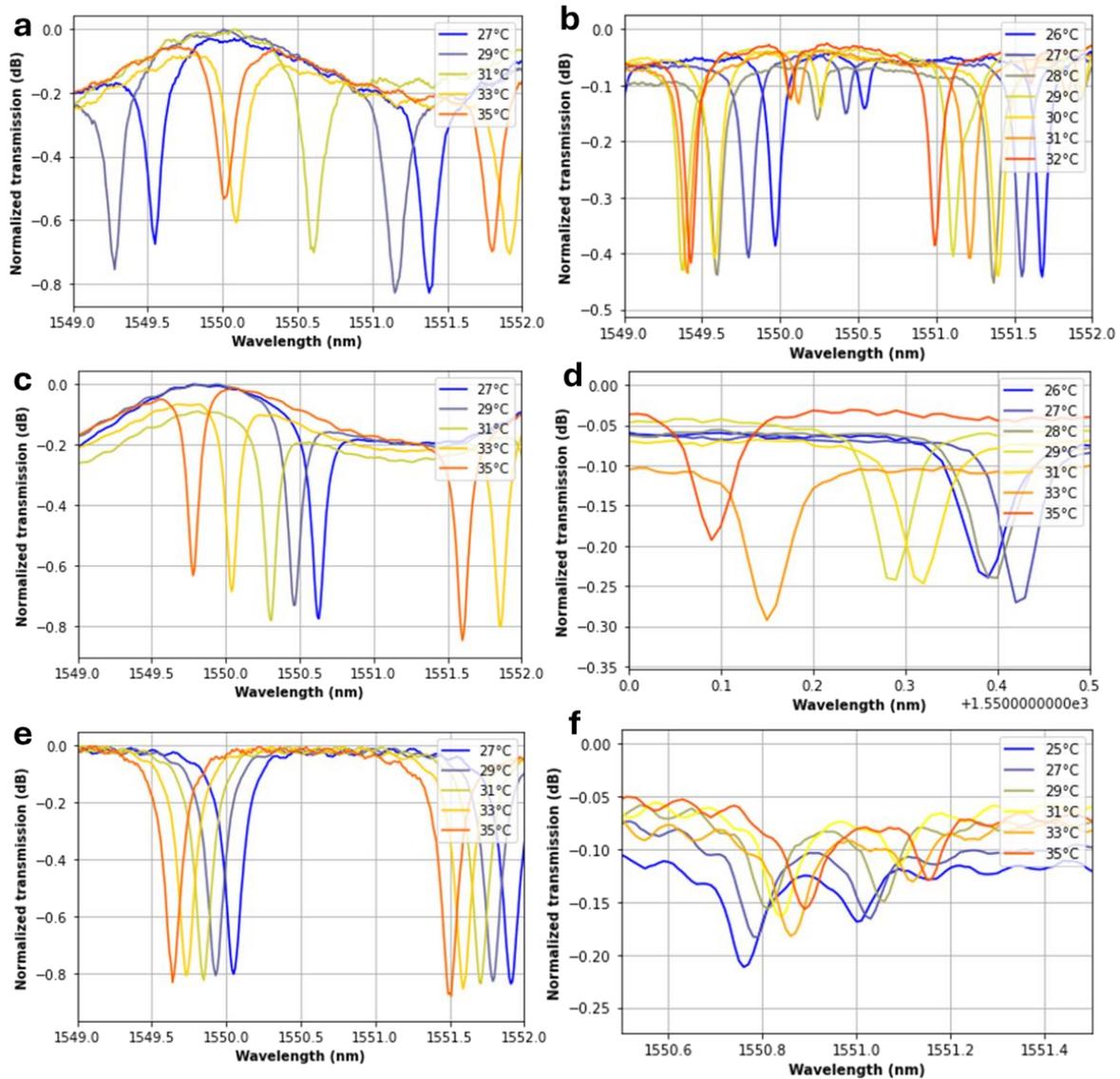

**Figure S14.** Wavelength shift as a function of temperature for SiN device using ICPCVD SiO$_2$ deposited at **a)** 30°C, **b)** 150°C, **c)** 300°C, **d)** 400°C, **e)** PMMA and **f)** PECVD SiO$_2$ deposited at 300°C

### 5.3. Silicon-On-Insulator devices

We deposited SOI optical devices using PECVD at temperatures of 200°C and 300°C and ICPCVD at a temperature of 75°C (**fig.S15**) and the wavelength spectra as a function of temperature is shown in **fig.S15**. We also deposited silicon dioxide via ICPCVD at 300°C and chamber pressure of 8 mTorr resulting in a thermal tunability of 5.5 pm/°C. The wavelength spectra as a function of the temperature and the corresponding fitting for the two resonances separated one free spectral range is shown in **fig.S16**.

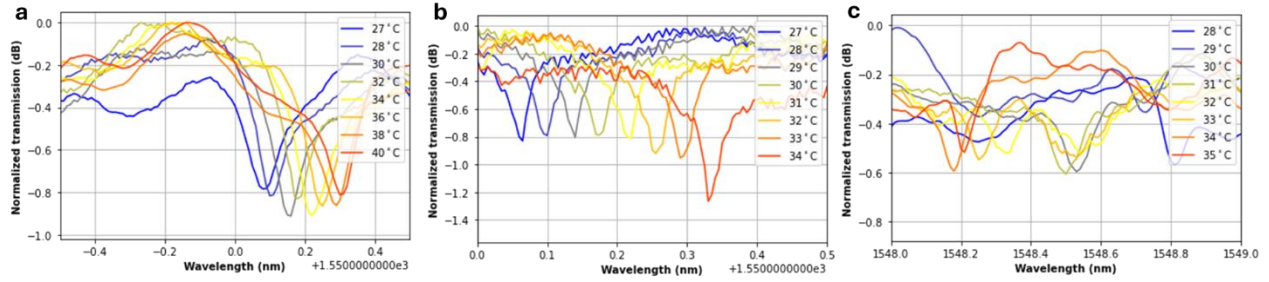

**Figure S15.** Wavelength shift as a function of temperature for an SOI device with SiO$_2$ claddings deposited using **a)** PECVD at 200°C, **b)** PECVD at 300°C and **c)** ICPCVD at 75°C.

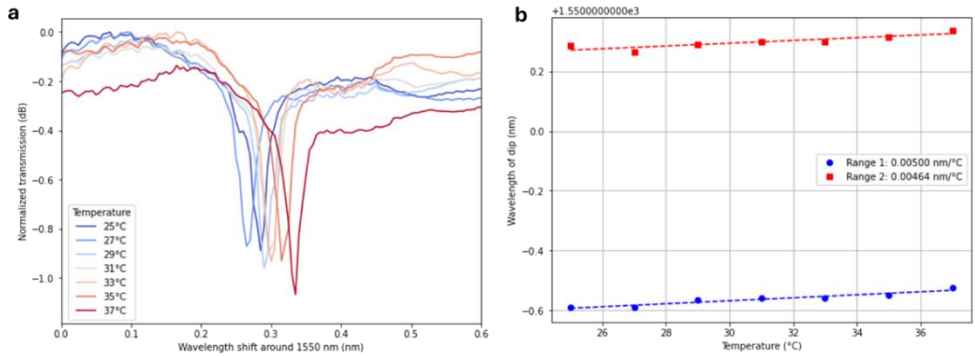

**Figure S16. a,** Transmission spectra for a Silicon-On-Insulator device with SiO$_2$ cladding deposited via ICPCVD at 300°C and chamber pressure of 8 mTorr and **b,** corresponding fitting of the two dips separated by a free spectral range.

## 6. Coupled-Resonator Optical Waveguide (CROW) devices

We fabricated one sample with two CROW devices using positive and negative claddings and metal micro-heaters. We measured this devices by sweeping the voltage and recorded the spectra.

### 6.1. CROW device 1

In the same configuration as the device shown in the main manuscript, we measured another device connected in parallel as depicted in **fig.S17a** in a voltage range of 0V to 12V in steps of 0.5V. **Fig.S17b** shows that the resonance condition can be achieved at a voltage of 9.5V. **Fig.S17c** shows wavelength spectra taken at different voltages of 0V, 7.5V and 9.5V (resonance condition).

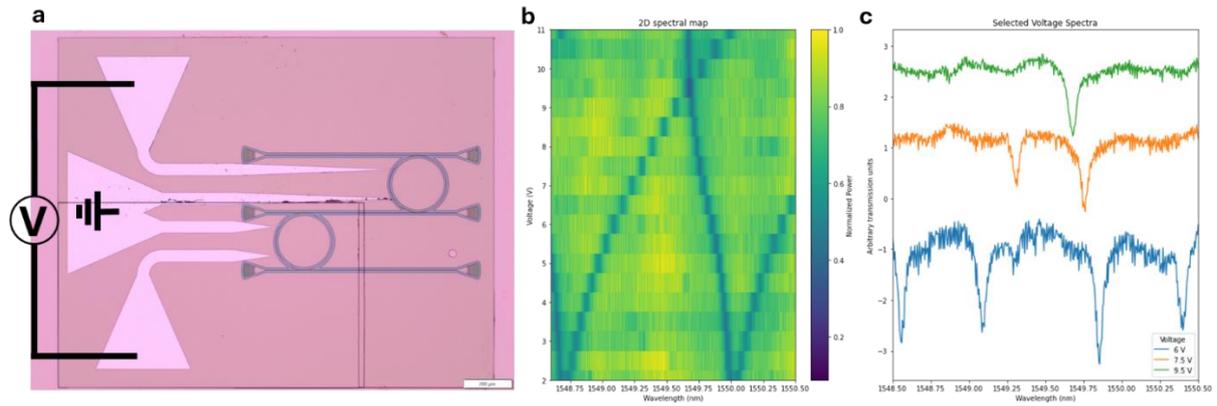

**Figure S17. a,** Optical microscope image of device 1 with two ring resonators connected with a middle waveguide to track the resonance dips in parallel connection configuration. **b,** Wavelength intensity spectra as a function of voltage applied and **c,** Wavelength spectra for 6V (blue), 7.5V (orange) and 9.5V (green).

## 6.2. CROW device 2

For the same device as the one shown in the manuscript (**fig.3b-c**) and in the same configuration (heaters connected in parallel), we did a coarse scan of the voltage in steps of 1 V from 0 to 13 V. A 2D mapping of the transmitted intensity as a function of the wavelength is shown in **fig.S18** together with the specific spectra taken at different voltages of 0V, 3V and 6V.

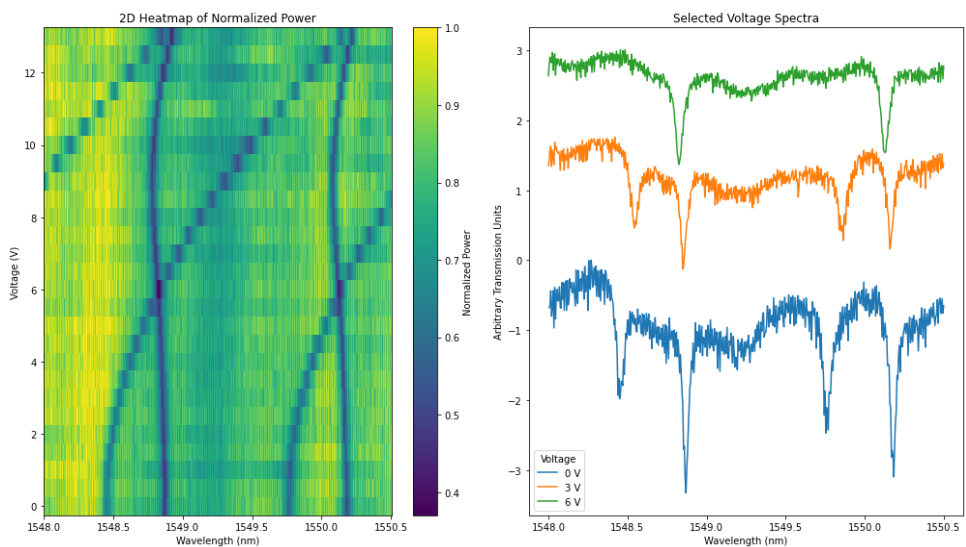

**Figure S18.** Transmission intensity as function of the wavelength spectra for different applied voltages for the device in fig.3b-c of the main manuscript and wavelength spectra for voltages 0V, 3V and 6V taken from the 2D map.

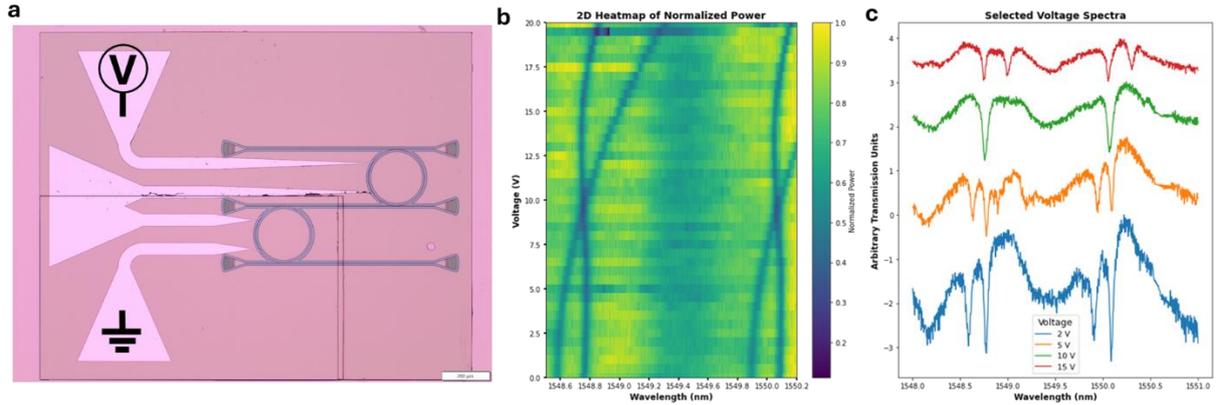

**Figure S19. a,** Optical microscope image of device 2 (shown in the main article fig.3b-c) connected in series with two ring resonators with a middle waveguide to track the resonance dips. **b,** Wavelength intensity spectra as a function of the volage applied and **c,** Wavelength spectra for 2V (blue), 5V (orange), 10V (green) and 15V (red).

## 7. Thermal crosstalk

In this section we show the corresponding spectra taken at different voltages (from 0V to 10V) to characterize the thermal crosstalk between devices using continuous PECVD (**fig.S20**) and ICPCVD (**fig.S21**) claddings as well as cladding deposited using lift-off for thermal isolation (**fig.S22**). Ring A depicts the device where the heater is applied while Ring B is the device that shifts due to thermal crosstalk. We also show an SEM image of the region between optical devices after performing ICPCVD lift-off of the cladding (**fig.S23**).

### 7.1. PECVD Continuous films

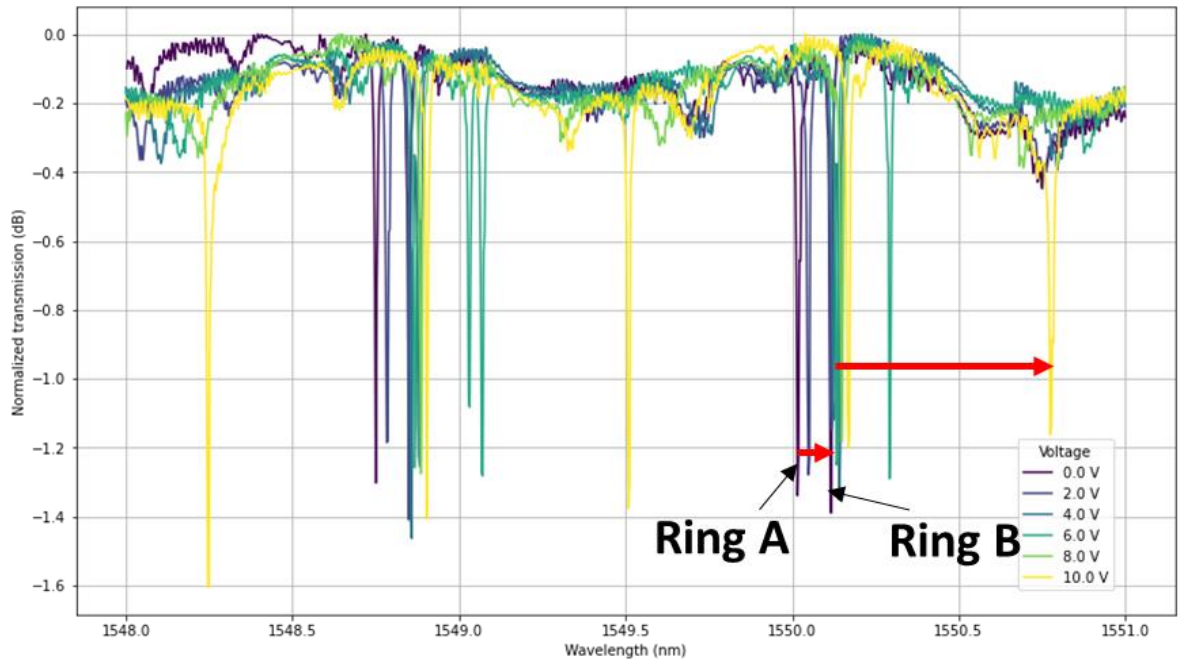

**Figure S20.** Spectra at different voltages for the two ring resonators fabricated using continuous PECVD cladding

## 7.2. ICPCVD Continuous film

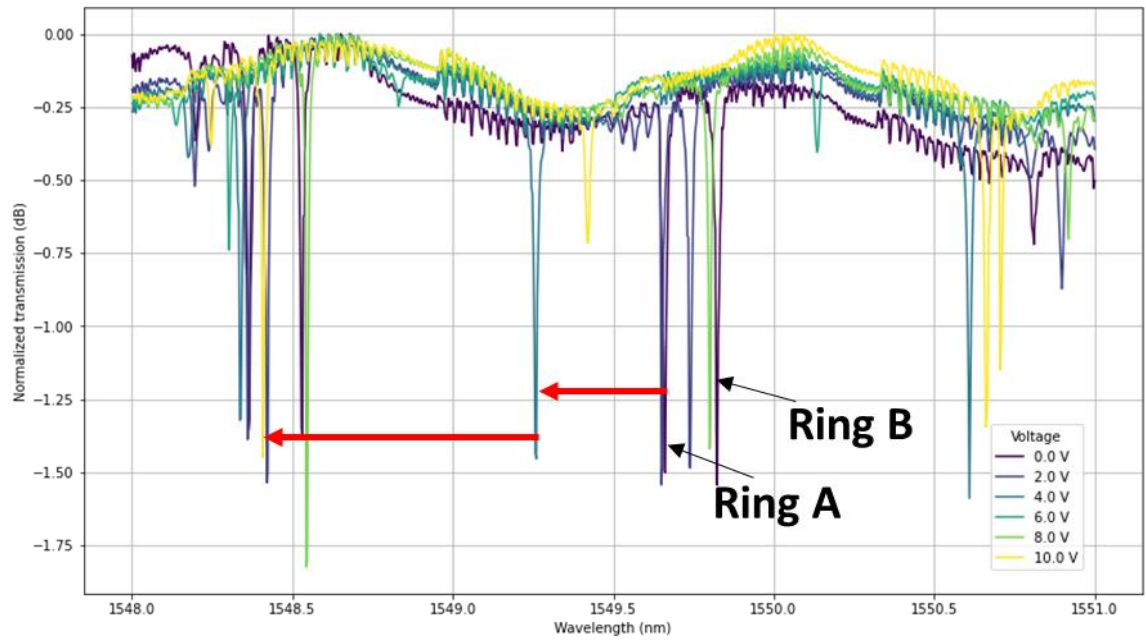

**Figure S21.** Spectra at different voltages for the two ring resonators fabricated using continuous ICPCVD cladding.

## 7.3. ICPCVD lift-off cladding

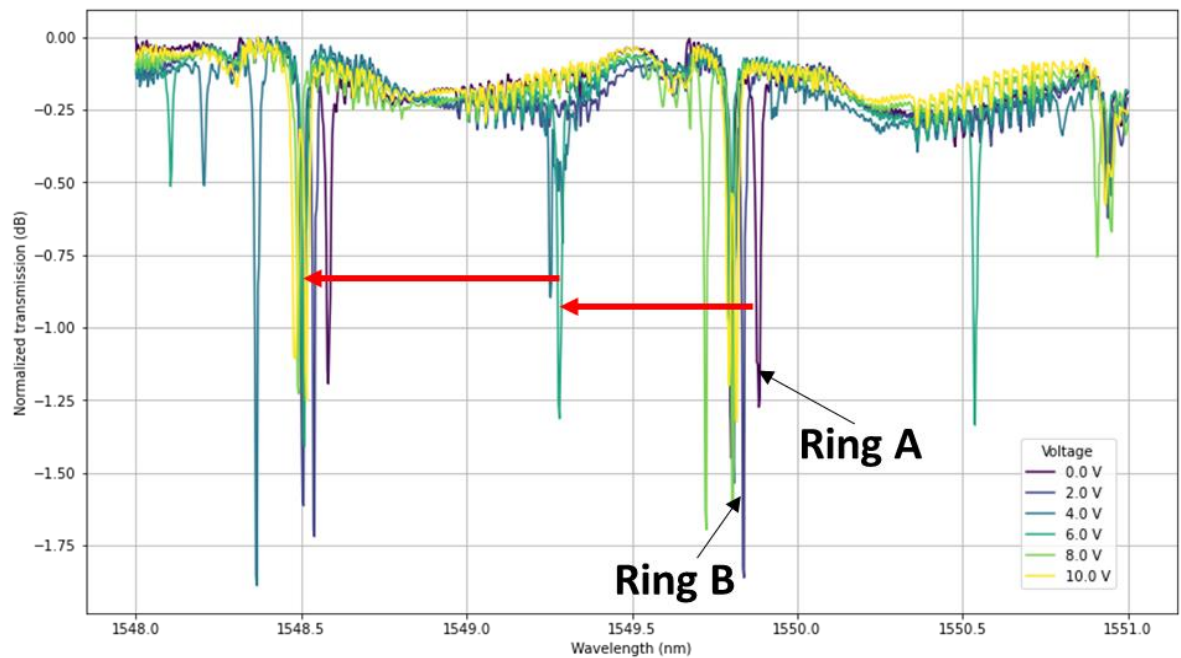

**Figure S22.** Spectra at different voltages for the two ring resonators fabricated using the lift-off approach.

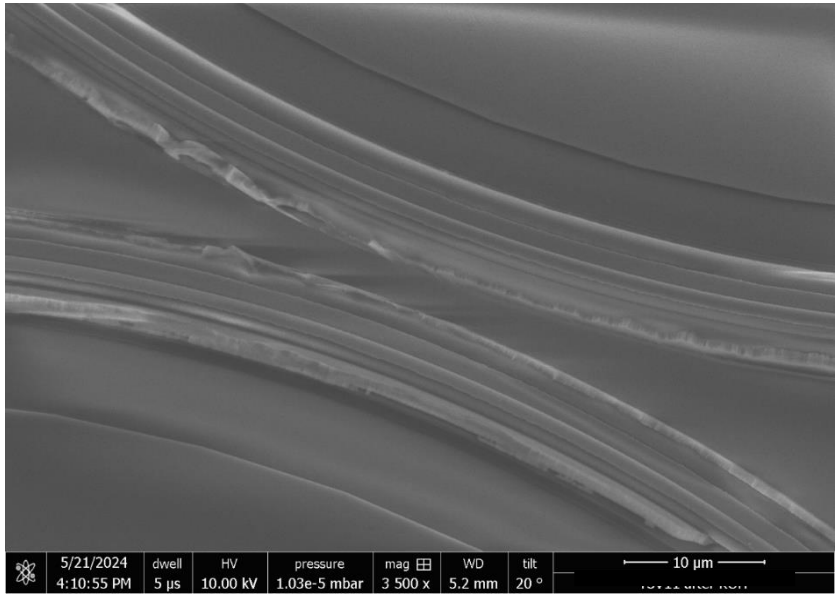

**Figure S23.** Scanning Electron Microscope imaging of the resulting cladding after lift-off in the region between two optical ring resonators.